\documentclass[aps,prd,reprint,showpacs,nofootinbib,superscriptaddress]{revtex4-1}
\usepackage{centernot}
\usepackage{relsize}
\usepackage{epsf}
\usepackage{epsfig}
\usepackage{graphics}
\usepackage{graphicx}
\usepackage{amssymb}
\usepackage{color}
\usepackage{multirow}
\usepackage{amsmath}
\usepackage{bigstrut}
\usepackage{blindtext}

\begin{document}


\title {\bf  Relativistic MOND from Modified Energetics}


\author{Durmu{\c s} Ali Demir}%
 \email{demir@physics.iztech.edu.tr}
\author{Canan Nurhan Karahan }%
\email{cananduzturk@iyte.edu.tr}
\affiliation{Department of Physics, {\.I}zmir Institute of Technology, TR35430, {\.I}zmir, Turkey}

\begin{abstract}
We begin to investigate the question of what modifications in energy-momentum
tensor can yield correct MOND regime. As a starting study, we refrain from
insisting on an action principle and focus exclusively on the equations of
motion. The present work, despite the absence of an explicit action functional, can be
regarded to extend Milgrom's modified inertia approach to relativistic domain.
Our results show that a proper MOND limit arises if energy-momentum tensor
is modified to involve determinant of the metric tensor in reference to the flat metric,
where the latter is dynamically generated as in gravitational Higgs mechanism. This
modified energy-momentum tensor is conserved in both Newtonian and MONDian regimes.
\end{abstract}

\maketitle
\section{Introduction}
Observations of several decades, ranging from the initial measurements by Oort (see the discussion in \cite{ref1}) to the
primal ones by Rubin \cite{ref2}, have shown that galaxies exhibit flat rotation curves, manifestly violating the Keplerian dynamics. This
universal anomalous dynamics has been interpreted in two distinct ways. The first, first proposed by Zwicky \cite{Zwicky} in 1933,
refers to Dark Matter (DM) hypothesis. According to the DM paradigm, there must be a distribution of non-shining  matter at the outer
skirts of galaxies to measure approximately constant velocities after particular distances from the centre of galaxies. The
DM hypothesis provides viable explanations not only for flat rotation curves but also for various cosmological and astronomical
observations describing different phases of the evolution of Universe. Several experimental groups have been searching for
DM particle by utilizing various detection methods (see the recent review volumes \cite{ref3}). So far, no signal of DM has
been observed.

The second interpretation, first proposed by Milgrom \cite{MOND123} in 1983, postulates that the observed flat rotation
curves result from modifications in the Newtonian laws of motion. In this approach, instead of adding unknown
ingredients to galactic matter, one exercises modifications in motion equations which dominate at the skirts of the
galaxies. To this end, Newton's law of motion $\vec{F} = m \vec{a}$ changes to
\begin{eqnarray}
\label{MOND}
\vec{F}=m\mu\left(\frac{a}{a_{0}}\right)\vec{a}
\end{eqnarray}
where $\vec{F}$ is the net force acting on the material point which has inertia $m$ and acceleration $\vec{a}$ (with $a^2=\vec{a}\cdot\vec{a}$). This
dynamical equation, structuring Milgrom's MOND theory \cite{MOND123}, is characterized by
the empirical function $\mu(a/a_0)$ where $a_{0}\simeq 1.2 \times 10^{-10} {\rm m s^{-2}}$ is a constant acceleration
scale for all galaxies \cite{horizon1}. It appears in (\ref{MOND}) as a critical acceleration scale set
galactically by the mass $M$ and radius $R$ of the galaxy as $(G_N M)/R^2 \simeq a_0$ and cosmologically
by the present-day value $H_0$ of the Hubble parameter as $(c H_0)/2\pi \simeq a_0$ \cite{horizon2}.

The heart of the MOND theory is the empirical function $\mu(a/a_0)$. There is yet no dynamical theory for it; however,
its asymptotic behavior is not difficult to guess
\begin{eqnarray}
\label{MOND-func}
\mu\left(x\right) \asymp \left\{
\begin{array}{l l}
       1 & \quad \mbox{if $x > 1 $}\\
       x & \quad \mbox{if $x < 1$}\\ \end{array} \right.
\end{eqnarray}
if all the successes of the Newtonian theory are to be maintained. Here $x$ does not need to be very large or small compared
to unity because $\mu(x)$ can attain its asymptotics even when $x$ is close to unity. For instance, the empirical form
\begin{eqnarray}
\label{MOND-funcp}
\mu(x) = \frac{x \left(\frac{3}{2}\right)^{\frac{2}{n^2}}}{\left(\frac{1}{1+x^n} + x^n\right)^{\frac{1}{n}}}
\end{eqnarray}
facilitates the asymptotics in (\ref{MOND-func}) almost independently of $x$ provided that $n$ is large.
Indeed, taking $n=50$ one finds $\mu(x) = 0.700$, 0.900, 0.986, 0.996, 1.000, 1.000 for
$x=0.7$, 0.9, 0.99, 1.01, 1.1, 2.0, respectively.

The behaviour in (\ref{MOND-func}) ensures that matter in the galaxy exhibits flat rotation curves
far away from the galactic center. Indeed, in the limit of small accelerations the equation of motion
(\ref{MOND}) takes the form
\begin{eqnarray}
\label{quad}
\vec{F} = m \frac{a\vec{a}}{a_{0}}
\end{eqnarray}
so that at large radii $R$ corresponding to outer skirts of the galaxy one finds not the Keplerian
law $|\vec{F}| = (m v^2)/R$ but $|\vec{F}| = (m v^4)/(a_0 R^2)$ which yields the constant speed
\begin{eqnarray}
\label{speed}
v^4= G_N M a_{0}
\end{eqnarray}
for $|\vec{F}| = (G_N m M)/R^2$. This relation accounts for the observed flat rotation curves \cite{ref2,FRC123}. The
constant speed (\ref{speed}) is the reason for and result from the whole idea of MOND. It depends crucially on
the behaviour of the empirical function (\ref{MOND-func}) at low accelerations.

The empirical MOND relation in (\ref{MOND}), supported by (\ref{MOND-func}) and (\ref{MOND-funcp}),  needs be formulated at a more fundamental level. In this regard, there arises
two different interpretations. In the first, after setting $\vec{a} = - \vec{\nabla} \phi_g$ with $\phi_g$ being the
gravitational potential, one formulates MOND as a modification in gravitational laws (see the reviews \cite{review}).
In this case, one is necessarily led to modified Newtonian gravity \cite{nr-teves} or General Relativity (GR) extended by geometrical
scalar and vector fields \cite{gr-teves, aether}. Besides, there are alternative approaches based on  $f(R)$ gravity \cite{fR},
bimetric gravity \cite{bimetric}, time foliation \cite{khronon}, nonlocal metric theories \cite{non-local-metric}, Galileons
\cite{galileon}, and Horava-Lifshitz gravity \cite{horava}. In general, modified gravity theories introduced to replace the DM
necessarily lead to MONDian structure.

In the second interpretation, one conceives the equation of motion (\ref{MOND}) as defining an acceleration-dependent
inertia $m(a) = m \mu({a}/{a_{0}})$. This approach, the modified inertia approach proposed in \cite{modif-inertia}, in
the non-relativistic limit, keeps gravitational laws unchanged yet lets in nonlinear kinetic terms. In this framework,
it is found that the kinetic term of the point mass involves all derivatives of acceleration \cite{modif-inertia,modif-inertia-2} yet
it is stable and respects causality \cite{review,non-local-metric}. In the present work, we pursue this modified inertia viewpoint to
generalize it to general-relativistic domain. The experience from non-relativistic study \cite{modif-inertia} ensures
that forming an action functional must be difficult, if not impossible, in the relativistic domain. We thus focus exclusively
on the equations of motion without specifying an action principle to derive them.

\section{Modified Energetics}
As the beginning phase of a study programme aiming at finding dynamical alternatives to modified gravity models of relativistic MOND \cite{review},
in this section we study gravitational field equations where MOND phase is understood as changes in matter energy-momentum tensor. This
approach, aiming at carrying Milgrom's modified inertia approach \cite{modif-inertia} into relativistic domain at the level of
equations of motion, is based on the matter energy-momentum tensor $T_{\mu\nu}^{(N)}$ in Newtonian domain and exploits its expected
non-conservation in the MOND regime to derive MONDian dynamics in an empirical way. Having a complete knowledge of the interactions
of matter, its energy-momentum tensor $T_{\mu\nu}^{(N)}$ (with energy density ${T}^N_{0 0}$, pressure ${T}^N_{i i}$, momentum density
${T}^N_{0 i}$ and shear stress  ${T}^N_{i j}$) is strictly conserved in the Newtonian regime. However, the same $T_{\mu\nu}^{(N)}$  is not conserved
in the MONDian regime because matter develops extra interactions even if one is not able to know them explicitly. Those extra
interactions generalize $T_{\mu\nu}^{(N)}$ to a conserved energy-momentum tensor $T_{\mu\nu}$ which can be approached
only empirically in the absence of a complete dynamical model (see \cite{higher-curve-eom} for a similar approach to
modified gravity framework for MOND). We now give an empirical implementation of this dynamical picture starting with Einstein field equations
\begin{eqnarray}
\label{EFE}
G_{\mu \nu} = 8\pi G_{N} {{T}}_{\mu \nu}
\end{eqnarray}
in which ${T}_{\mu \nu}$ is the conserved energy-momentum tensor of matter at all acceleration scales ranging from ${a}=0$ to ${a}=\infty$.
In general, $T_{\mu\nu}$ is conserved on the equations of motion, and these equations necessarily encode the novel interactions of
matter responsible for the MOND. However, those new interactions are not known and our knowledge of $T_{\mu\nu}$ is incomplete;
we are able to know it only when $a>a_0$ for which it equals $T^{(N)}_{\mu\nu}$. Consequently, on an empirical basis we write for $T_{\mu\nu}$
\begin{eqnarray}
\label{en-mom-tot}
{T}_{\mu \nu}= \mu\left({\mathlarger{\mathlarger{\mathfrak{a}}}}\right)\left[T^{(N)}_{\mu \nu}- {Q} g_{\mu\nu}\right] + {Q} g_{\mu\nu}
 \end{eqnarray}
where $\mu(x)$ is the MOND function in (\ref{MOND-func}), $Q$ is a scalar, and ${\mathlarger{\mathlarger{\mathfrak{a}}}}$ is
yet another scalar which is to be judiciously  constructed to have the empirical limit
\begin{eqnarray}
\label{NR-limit}
{\mathlarger{\mathlarger{\mathfrak{a}}}} \xrightarrow{v \ll c} {\mathlarger{\mathlarger{\mathfrak{a}}}}_{NR} = \frac{a}{a_0}
\end{eqnarray}
at non-relativistic energies. This correspondence between the relativistic (${\mathlarger{\mathlarger{\mathfrak{a}}}}$) and
non-relativistic ($a$) regimes is crucial for the empirical structure in (\ref{en-mom-tot}) to give a consistent framework.

Physically, the grand energy-momentum tensor ${T}_{\mu \nu}$ must correctly reproduce the Newtonian and MONDian regimes. This
is analyzed case by case in Table \ref{table1} as a function of the divergence of $T^{(N)}_{\mu\nu}$. As suggested by the
table, underlying dynamics can be revealed after a proper understanding of ${T}_{\mu\nu}$ and this
requires $T^{(N)}_{\mu\nu}$, $\mathfrak{a}(T)$ and ${Q}(T)$ to be constructed in detail. We detail these physical
variables in the three consecutive subsections that follow.

\begin{table*}[t]
\begin{center}
    \begin{tabular}{ | l | l | l | p{6cm} |}
    \hline
    Acceleration & MOND Function & Energy-Momentum Tensor & Matter Dynamics \\ \hline
    ${\mathlarger{\mathlarger{\mathfrak{a}}}} \gtrsim 1$ & $\mu\left({\mathlarger{\mathlarger{\mathfrak{a}}}}\right) \simeq 1$ & $\begin{array}{l} {T}_{\mu \nu}\simeq T_{\mu \nu}^{(N)}\\\\
\left(\nabla^{\mu} {T}_{\mu \nu} = 0\ \text{hence}\ \nabla^{\mu} T^{(N)}_{\mu \nu} = 0\right)\end{array}$ & This is `Newtonian regime'. Acceleration
of matter is above $a_0$ and $\mu\left({\mathlarger{\mathlarger{\mathfrak{a}}}}\right)$ ensures ${{T}}_{\mu\nu} \simeq T^{(N)}_{\mu\nu}$ so that $T^{(N)}_{\mu\nu}$ is symmetric and divergence-free
($\nabla^{\mu} T^{(N)}_{\mu\nu} = 0$) in agreement with (\ref{EFE}). In Newtonian regime thus $T^{(N)}_{\mu\nu}$ qualifies as the known conserved
energy-momentum tensor of matter. \\ \hline
    ${\mathlarger{\mathlarger{\mathfrak{a}}}} \lesssim 1$ & $\mu\left({\mathlarger{\mathlarger{\mathfrak{a}}}}\right) \simeq {\mathlarger{\mathlarger{\mathfrak{a}}}}$ &
$\begin{array}{l} {T}_{\mu \nu}\centernot\simeq T^{(N)}_{\mu \nu}\\\\
\left(\nabla^{\mu} {T}_{\mu \nu} = 0\ \text{yet}\ \nabla^{\mu} T^{(N)}_{\mu \nu} \neq 0 \right)\end{array}$& This is `MONDian regime'.
 Acceleration of matter is below $a_0$ and $\mu\left({\mathlarger{\mathlarger{\mathfrak{a}}}}\right)$ leads to ${{T}}_{\mu\nu} \centernot\simeq T^{(N)}_{\mu\nu}$ so that $T^{(N)}_{\mu\nu}$ is
symmetric yet not divergence-free ($\nabla^{\mu} T^{(N)}_{\mu\nu} \neq 0$). In MOND regime thus it is ${T}_{\mu\nu}$
not $T^{(N)}_{\mu\nu}$ which qualifies as the conserved energy-momentum tensor of matter. In this small acceleration regime, matter
develops novel interactions that make $\nabla^{\mu} T^{(N)}_{\mu\nu} \neq 0$ yet the scalars ${\mathlarger{\mathlarger{\mathfrak{a}}}}$ and ${Q}$
help ${T}_{\mu\nu}$ be conserved and give the observed flat rotation curves. \\ \hline
\end{tabular}
\end{center}

\caption{The acceleration dependence of the energy-momentum tensor ${T}_{\mu\nu}$ of matter. In general, ${\mathlarger{\mathlarger{\mathfrak{a}}}} = {\mathlarger{\mathlarger{\mathfrak{a}}}}(T^{(N)})$
and ${Q} = {Q}(T^{(N)})$ are functions of the energy-momentum tensor $T^{(N)}_{\mu\nu}$. These scalars
take appropriate values for Newtonian ($T^{(N)}_{\mu\nu}$ is conserved) and MONDian ($T^{(N)}_{\mu\nu}$ is not conserved) regimes.
Namely, matter develops novel interactions (such as the higher-derivative kinetic terms,
determined in \cite{modif-inertia} in the non-relativistic regime) at small accelerations and its known
energy-momentum tensor $T^{(N)}_{\mu\nu}$ starts exhibiting non-conservation properties.}\label{table1}
\end{table*}

\subsection{Physical Properties of $T^{(N)}_{\mu\nu}$}
It has been emphasized previously, specifically in Table \ref{table1}, that $T^{(N)}_{\mu\nu}$ has the same form as the energy-momentum
tensor of matter in Newtonian regime yet it does not qualify as true energy-momentum tensor in the MOND regime simply because its
conservation is spoiled by novel interactions of matter that arise at accelerations below $a_0$. The higher-derivative self interactions
studied in \cite{modif-inertia,modif-inertia-2} form a concrete example of such effects. Let us consider, as an illustrative example,
dust (pressureless matter having only energy density in the comoving frame) for which
\begin{eqnarray}
\label{en-mom-dust}
T^{(N)}_{\mu\nu} = \rho u_{\mu} u_{\nu}
\end{eqnarray}
where $\rho$ and $u_{\mu}$ are energy density and velocity, respectively. ( One recalls that
$T^{(N)}_{\mu\nu} = \int d \tau \rho u_{\mu} u_{\nu}$ for a relativistic particle with
trajectory $y_{\mu}(\tau)$ and energy density $\rho = m c^2 \delta^{4}(x-y(\tau))$.) It is
divergence-free, $\nabla^{\mu} T^{(N)}_{\mu \nu} = 0$, because densities and flows of
dust are all conserved. However, this conservation property holds only in normal
circumstances where Newtonian laws of motion are valid. In MONDian regime, where dust develops
higher-derivative kinetic interactions for instance, conservation breaks down, $\nabla^{\mu} T^{(N)}_{\mu \nu} \neq 0$.
On dimensional grounds, it is likely to have structures of the form
\begin{eqnarray}
\label{non-conserve}
\nabla^{\mu} T^{(N)}_{\mu \nu} \sim \rho a_0 u_{\nu}
\end{eqnarray}
in addition to terms involving derivatives of acceleration. In the absence of an invariant
action (like the non-relativistic model in \cite{modif-inertia}), this non-conservation can
be understood neither in origin nor in structure ($\rho a_0 u_{\nu}$ in (\ref{non-conserve}) is
just an example). Therefore, our goal is not to construct a model of the
non-conservation of $T^{(N)}_{\mu\nu}$ but to determine its consequences for structures and
dynamics of ${\mathlarger{\mathlarger{\mathfrak{a}}}}$ and ${Q}$.

\subsection{Physical Properties of the Acceleration Scalar ${\mathlarger{\mathlarger{\mathfrak{a}}}}$}
The acceleration scalar ${\mathlarger{\mathlarger{\mathfrak{a}}}}$, which must have the non-relativistic
limit ${\mathlarger{\mathlarger{\mathfrak{a}}}}_{NR}$ given in (\ref{NR-limit}), must be constructed
judiciously to correctly cover the Newtonian and MONDian regimes. Hence, besides the crucial
relation (\ref{NR-limit}), it must have the following properties.
\begin{enumerate}
\item By our construction shown in Table \ref{table1}, ${\mathlarger{\mathlarger{\mathfrak{a}}}}$  must vary with the
divergence of $T^{(N)}_{\mu\nu}$ as
\begin{eqnarray}
\label{en-mom-cond}
\begin{array}{l}
{\mathlarger{\mathlarger{\mathfrak{a}}}} > 1\;\; \text{if}\;\;\; \nabla^{\mu} T^{(N)}_{\mu \nu} = 0\\ \\
{\mathlarger{\mathlarger{\mathfrak{a}}}} < 1\;\; \text{if}\;\;\; \nabla^{\mu} T^{(N)}_{\mu \nu} \neq 0
\end{array}
\end{eqnarray}
while $\nabla^{\mu} {{T}}_{\mu \nu} = 0$ in both cases.

\item Being a scalar field, ${\mathlarger{\mathlarger{\mathfrak{a}}}}$ involves contractions of the divergences of
$T^{(N)}_{\mu\nu}$. This necessarily brings in the gravitational acceleration $\vec{\nabla} \phi_g$ through the
gravitational potential $\phi_g = -1-g_{00}$ arising in the Newtonian limit of the metric tensor $g_{\mu\nu}$. However,
presence of $\vec{\nabla}\phi_g$ must be prohibited for ${\mathlarger{\mathlarger{\mathfrak{a}}}}$ to yield the
kinetic acceleration in (\ref{NR-limit}). It is easy to see that this cannot be accomplished
without using an independent source of $\phi_g$ and the most natural source as such is the determinant
$g={\mbox{Det}}(g_{\mu\nu})$ of the metric tensor. However, being a scalar density rather than a
scalar, $g$ cannot appear in ${\mathlarger{\mathlarger{\mathfrak{a}}}}$ by itself; it must be divided by another
scalar density to achieve covariance. This other scalar density necessitates a new metric
$\overline{g}_{\mu\nu}$, and  naturally leads to a bi-metrical picture (whose relevance for MOND has been
discussed in \cite{bimetric}). Then, acceleration scalar possess the functional form
\begin{eqnarray}
\label{a-func}
{\mathlarger{\mathlarger{\mathfrak{a}}}} = {\mathlarger{\mathlarger{\mathfrak{a}}}}\left(a_0, \nabla^{\mu} T^{(N)}_{\mu \nu}, T^{(N)}_{\mu\nu}, g_{\mu\nu},
g_{\mu\nu} \overline{g}^{\mu\nu}, {g}/{\overline{g}}\right)
\end{eqnarray}
where $\overline{g} = {\mbox{Det}}(\overline{g}_{\mu\nu})$ arises as an additional variable to be dynamically determined.
\end{enumerate}
These two points plus (\ref{NR-limit}) must be taken into account in formulating ${\mathlarger{\mathlarger{\mathfrak{a}}}}$.
However, the formulation process  becomes utterly incomplete unless the additional metric $\overline{g}_{\mu\nu}$ is demystified.
In the two subsections that follow, we first study $\overline{g}_{\mu\nu}$ and then construct a model of ${\mathlarger{\mathlarger{\mathfrak{a}}}}$.
\vspace{0,1 cm}
\subsubsection{Construction of $\overline{g}_{\mu\nu}$}
The second metric tensor $\overline{g}_{\mu\nu}$, required to eliminate the gravitational acceleration $\vec{\nabla} \phi_g$ from
the acceleration scalar ${\mathlarger{\mathlarger{\mathfrak{a}}}}$, can be ascribed different structures depending on the
underlying dynamics. For instance, one may consider identifying it with $T^{(N)}_{\mu\nu}$ itself  but this attempt fails
because its determinant vanishes in the case of dust (see equation (\ref{en-mom-dust}) above)). Alternatively, one may take $\overline{g}_{\mu\nu}$
as a second metric tensor with its own curvature and dynamics but this setup, as was already elaborated by Milgrom in (\cite{bimetric}) (see also \cite{woodard}),
gives a modified gravity theory for MOND. This and other possible modified gravity models fall outside the scope of the present
work because the goal here is to develop a dynamical approach to relativistic MOND similar
in philosophy to Milgrom's modified inertia approach \cite{modif-inertia}.

Our approach to $\overline{g}_{\mu\nu}$ is dynamical rather than geometrical. In other words, the dynamics underlying
the asymptotics in Table \ref{table1} and structures in (\ref{a-func}) proceed with not only $T^{(N)}_{\mu\nu}$ but also
$\overline{g}_{\mu\nu}$. Thus, $\overline{g}_{\mu\nu}$ is a low-acceleration dynamical field, maybe
one of many as such, which facilitates the MOND regime. In modeling the dynamics, we interpret the
coupling $g^{\mu\nu} \overline{g}_{\mu\nu}$ between the two metrics as the kinetic term of
four real scalars $\phi^{m}$ ($m=0,\dots,3$), and construct the defining relation
\begin{eqnarray}
\label{metric-bar}
\overline{g}_{\mu\nu} = \frac{1}{M^4} \eta_{m n } \partial_{\mu}\phi^m \partial_{\nu}\phi^n
\end{eqnarray}
where $\eta_{m n}$ is the flat Minkowski metric, and hence, scalar spectrum contains a
a ghosty (negative kinetic term) mode. We assume that $\phi^{m}$ develop the nontrivial backgrounds
\begin{eqnarray}
\label{vacua-metrics}
\langle \overline{g}_{\mu\nu}\rangle  = \left\{\begin{array}{l} 0\;\;\;\;\; \mbox{if}\;\;\; \langle \phi^m \rangle = 0\\
\eta_{\mu\nu} \;\; \mbox{if}\;\;\; \langle \phi^m \rangle = M^2 x^m \end{array} \right.
\end{eqnarray}
depending on whether the diffeomorphism invariance is exact ($\langle \phi^m \rangle = 0$) or spontaneously
broken ($\langle \phi^m \rangle = M^2 x^a$) in the vacuum state governed by the vacuum expectation value
$\langle \phi^a \rangle$ of the scalars. Here, the scale $M$ is around $a_0$. The dynamics leading to (\ref{vacua-metrics}) can be known only
in a setting where all interactions of matter and extra fields like $\phi^{a}$ are specified. The diffeomorphism-breaking
vacuum here sets the flat Minkowski metric $\eta_{\mu\nu}$ as the background metric about which $g_{\mu\nu}$ can be expanded
in a perturbation series.

This induction mechanism is similar to what happens in gravitational Higgs mechanism \cite{higgs-mech, demirpak} in which
a second metric tensor $\overline{g}_{\mu\nu}$ is needed for writing a sensible graviton mass term through the kinetic
term $g^{\mu\nu} \overline{g}_{\mu\nu}$ of scalars and through the ratio of the determinants $g/\overline{g}$. Nevertheless,
as was throughly analyzed in \cite{demirpak}, these two contributions, instead of adding, can cancel each other
to keep graviton massless, or equivalently, gravity unmodified. This does not mean that the metric tensors in
(\ref{vacua-metrics}) do not participate in other physical processes. Indeed, they can well generate our targeted
structures involving the gravitational acceleration $\vec{\nabla} \phi$. Consequently, we associate the metric tensors
in (\ref{vacua-metrics}) with the two phases of motion as
\begin{eqnarray}
\label{vacua-metrics-2}
\begin{array}{l} \langle \overline{g}_{\mu\nu} \rangle = 0\;\;\;\;\; \Longrightarrow\;\;\; \mbox{Newtonian regime}\\
\langle \overline{g}_{\mu\nu} \rangle = \eta_{\mu\nu}\;\; \Longrightarrow\;\;\;  \mbox{MONDian regime}\end{array}
\end{eqnarray}
keeping in mind that gravity is not necessarily massive. Indeed, the model of \cite{demirpak}
offers a wide parameter space to set $V_1^{\prime}(4) = 0$ in equation (26) and $\zeta V_1^{\prime}(4) = 0$ in equation
(27). Moreover, potential terms in equation (11) give enough freedom to realize
massless and massive gravity phases. Therefore, as will be proven below, the MOND regime can be
realized by using the metrics in (\ref{vacua-metrics}) without the necessity of modifying gravity.

\subsubsection{Construction of ${\mathlarger{\mathlarger{\mathfrak{a}}}}$}
Having fixed all the variables in (\ref{a-func}), we now start formulating the acceleration scalar ${\mathlarger{\mathlarger{\mathfrak{a}}}}$.
The kinetic term $g^{\mu\nu} \overline{g}_{\mu\nu}$ of scalars do not contribute to $\vec{\nabla} \phi_g$, and hence,
the argument of ${\mathlarger{\mathlarger{\mathfrak{a}}}}$ in (\ref{a-func}) represent the
optimal list of dynamical variables. Out of various possibilities, we consider for ${\mathlarger{\mathlarger{\mathfrak{a}}}}$ a simple structure
\begin{widetext}
\begin{eqnarray}
\label{relat}
{\mathlarger{\mathlarger{\mathfrak{a}}}}^2 a_0^2 \left(T^{(N)}\right)^2 = \nabla^{\alpha} {T^{(N)}}_{\alpha}^{\beta}\nabla^{\theta}T^{(N)}_{\theta \beta} +
{c_{1}} \left(T^{(N)}\right)^2 {\nabla_{\alpha}\left(\frac{g}{\overline{g}}\right)\nabla^{\alpha}\left(\frac{g}{\overline{g}}\right)}
+ {c_2} T^{(N)}  {\nabla^{\alpha}\left(\frac{g}{\overline{g}}\right)\nabla^{\theta}T^{(N)}_{\theta\alpha}}
\end{eqnarray}
\end{widetext}
where all indices are raised and lowered with $g_{\alpha\beta}$ so that $T^{(N)}=g^{\alpha\beta} T^{(N)}_{\alpha \beta}$ is
the trace of the matter energy-momentum tensor in Newtonian domain. Here, the dimensionless constants $c_{1,2}$ will
be fixed in the weak field limit by imposing (\ref{NR-limit}). The presence of the metric determinants in (\ref{relat})
is crucially important for MOND because gravitational acceleration $\vec{\nabla}\phi_g$ is generated by derivatives of
$g/\overline{g}$ (not $g^{\mu\nu} \overline{g}_{\mu\nu}$, for instance).

Having fixed its functional form in (\ref{relat}), we now start checking if ${\mathlarger{\mathlarger{\mathfrak{a}}}}$
satisfies its defining asymptotics in (\ref{MOND-func}) and Table \ref{table1}. This requires its evaluation
in the two vacua in (\ref{vacua-metrics}) since they correspond to the Newtonian and MONDian regimes
as indicated in (\ref{vacua-metrics-2}).
\begin{enumerate}
\item {\it $\langle \overline{g}_{\mu\nu}\rangle  = 0$ and $\nabla^{\mu} T^{(N)}_{\mu\nu} = 0$.}
In this vacuum, $\langle \overline{g} \rangle$ vanishes identically and, as follows from (\ref{relat}),
${\mathlarger{\mathlarger{\mathfrak{a}}}}$ becomes infinitely large thanks to the fact that $c_{1,2} > 0$,
as will be proven below. Now, having found ${\mathlarger{\mathlarger{\mathfrak{a}}}} > 1$, one gets
$\mu\left({\mathlarger{\mathlarger{\mathfrak{a}}}}\right) \simeq 1$ and this gives $T_{\mu\nu} \simeq T^{(N)}_{\mu\nu}$
from (\ref{en-mom-tot}). Thus, the Einstein field equations (\ref{EFE}) reduce to
\begin{eqnarray}
G_{\mu \nu} = 8\pi G_{N} T^{(N)}_{\mu \nu}
\end{eqnarray}
in which consistency of the Bianchi identity on $G_{\mu\nu}$ is maintained by the conservation of $T^{(N)}_{\mu\nu}$.
This conservation, $\nabla^{\mu} T^{(N)}_{\mu\nu} = 0$, gives the usual Newtonian equations for free-fall
\begin{eqnarray}
\label{free-fall}
\vec{a} = - \vec{\nabla} \phi_g
\end{eqnarray}
for dust distribution characterized by the energy-momentum tensor in (\ref{en-mom-dust}). Clearly, this equation holds
if the metric tensor takes the form
\begin{eqnarray}
\label{metric-newton}
g_{\mu\nu} = {\mbox{Diag.}}\left(-(1 + 2 \phi_g), 1, 1, 1 \right)_{\mu\nu}
\end{eqnarray}
as appropriate for the non-relativistic limit.

In conclusion, as conjectured in equation (\ref{vacua-metrics-2}), the minimum energy configuration $\langle \overline{g} \rangle$
gives rise to the Newtonian regime for motion. Small perturbations about this vacuum makes $\overline{g} \neq 0$ but this determinant
is expected to be sufficiently small to secure the Newtonian regime ${\mathlarger{\mathlarger{\mathfrak{a}}}} > 1$.

\item {\it $\langle \overline{g}_{\mu\nu}\rangle  = \eta_{\mu\nu}$ and $\nabla^{\mu} T^{(N)}_{\mu\nu} \neq 0$.}
In this vacuum, in the non-relativistic limit in which metric tensor is given by (\ref{metric-newton}),
the acceleration scalar defined in (\ref{relat}) becomes
\begin{eqnarray}
\label{nonrelat}
{\mathlarger{\mathlarger{\mathfrak{a}}}}_{NR}^2 &=&\frac{ \vec{a}\cdot\vec{a}}{a_{0}^{2}}+ \left(2-c_{2}\right) \frac{\vec{a}\cdot \vec{\nabla} \phi_g} {a_{0}^{2}}\nonumber\\
&+& \left(1-c_{2}+c_{1}\right) \frac{\vec{\nabla}\phi_g \cdot \vec{\nabla}\phi_g}{a_{0}^{2}}
\end{eqnarray}
for dust whose energy-momentum tensor is given partly by (\ref{en-mom-dust}) and partly by extra interactions occurring in low-acceleration
regime. It is due to this alleged extra piece that $T^{(N)}_{\mu\nu}$ in (\ref{en-mom-dust}) satisfies $\nabla^{\mu} T^{(N)}_{\mu\nu} \neq 0$.

It is clear that, the acceleration scalar exhibits correct non-relativistic limit if
\begin{eqnarray}
c_1 = 1\;,\;\; c_2 = 2
\end{eqnarray}\\
because then the last two terms of (\ref{nonrelat}) drop out to enable the required limit in (\ref{NR-limit}). Thus, the construct in
(\ref{relat}) for ${\mathlarger{\mathlarger{\mathfrak{a}}}}$ does indeed reduce to the acceleration of the point mass rather than the
gravitational acceleration $-\vec{\nabla}\phi_g$. The non-relativistic result in (\ref{nonrelat}), which holds for
${\mathlarger{\mathlarger{\mathfrak{a}}}} < 1$ or equivalently $a < a_0$, entails $\mu\left({\mathlarger{\mathlarger{\mathfrak{a}}}}\right) \simeq {\mathlarger{\mathlarger{\mathfrak{a}}}}$
so that Einstein field equations (\ref{EFE}) takes the form
\begin{eqnarray}
\label{einstein-mond}
G_{\mu \nu} = 8 \pi G_{N}\left\{ {\mathlarger{\mathlarger{\mathfrak{a}}}} \left[T_{N\, \mu \nu}- Q g_{\mu \nu}\right] + Q g_{\mu \nu}\right\}
\end{eqnarray}
where the scalar field $Q$ is to be chosen judiciously to make the right-hand side to have vanishing divergence. This constraint, ensuring
conservation of $T_{\mu\nu}$, can be difficult to satisfy if $Q$ does not involve $T^{(N}_{\mu\nu}$ and $g/\overline{g}$. As a plausible structure, we set
\begin{eqnarray}
\label{Q-scal}
Q = \frac{g\,T^{(N)}}{\overline{g}}
\end{eqnarray}
where one can of course consider alternative structures giving similar results in the non-relativistic limit. In $\langle \overline{g}_{\mu\nu}\rangle  \!= \!\eta_{\mu\nu}$ vacuum,
in the non-relativistic limit, conservation of $T_{\mu\nu}$ gives
\begin{widetext}
\begin{eqnarray}
\label{MOND-1}
\nabla_{\mu}\left(\frac{{a}}{a_{0}}\right)\left\{ \rho u^{\mu}u^{j}-\rho \left(\frac{g}{\overline{g}}\right)g^{\mu j}
\right\} &=& -\frac{{a}}{a_{0}}\left\{\rho \left(a^{j}+\nabla^{j}\phi_g \right)-\rho \nabla^{j}\phi_g- (1 + 2 \phi_g) \nabla_{\mu}\rho g^{\mu j}\right\}\nonumber \\
&-&\rho \nabla^{j}\phi_g- (1 + 2 \phi_g) \nabla_{\mu}\rho g^{\mu j}
\end{eqnarray}
\end{widetext}
where the metric tensor is given by (\ref{metric-newton}). This differential equation is too involved to suggest the MOND dynamics. Nevertheless, a closer
look reveals that, if {\it (i)} energy density $\rho$ varies slowly in space ($|\vec{\nabla}\rho| \ll \rho |\vec{\nabla}\phi_g|$) and if
{\it (ii)} acceleration $\vec{a}$ varies slowly both in space and time ($|{\nabla}_{\mu} \rho| \ll \rho |{\nabla}_{\mu}\phi_g|$) then one gets from (\ref{MOND-1})
\begin{eqnarray}
\label{mond-eom}
\frac{ a \vec{a}}{a_0} = - \vec{\nabla}\phi_g
\end{eqnarray}
which is the desired MOND relation given in equation (\ref{quad}).

\item {\it Non-Conservation of $T_{N}^{\mu\nu}$.}
Having obtained motion equations in the two regimes of $\overline{g}_{\mu\nu}$, we now turn to a discussion of the non-conservation of $T_{N}^{\mu\nu}$.
In view of the discussions summarized in Table \ref{table1}, the energy-momentum tensor $T_{\mu\nu}$, introduced in (\ref{EFE}) and defined in (\ref{en-mom-tot}),
is always conserved. This is necessary for the consistency of the gravitational field equations (\ref{EFE}). The $T_{N}^{\mu\nu}$ tensor, however, is conserved only
in the Newtonian regime. To see how these conservation features hold, it proves useful to examine the divergence of $T_{N}^{\mu\nu}$
\begin{eqnarray}
\nabla_{\mu} T_{N}^{\mu\nu} = f_{N}^{\nu}
\end{eqnarray}
where
\begin{eqnarray}
\label{force}
f_{N}^{\nu} &=& - [\nabla_{\alpha}\ln{\mu\left({\mathlarger{\mathlarger{\mathfrak{a}}}}\right)}] \left( T_{N}^{\alpha\nu} - Q g^{\alpha\nu}\right) \nonumber\\
&+& \left(1 - \frac{1}{\mu\left({\mathlarger{\mathlarger{\mathfrak{a}}}}\right)}\right) \nabla^{\nu}Q
\end{eqnarray}
as follows from (\ref{EFE}) with (\ref{en-mom-tot}). It is obvious that, in the Newtonian regime, $\mu\left({\mathlarger{\mathlarger{\mathfrak{a}}}}\right) \rightarrow 1$ and $f_{N}^{\nu}$ vanishes
identically to ensure conservation of $T_{N}^{\mu\nu}$. In MONDian regime, however, $\mu\left({\mathlarger{\mathlarger{\mathfrak{a}}}}\right) \rightarrow {\mathlarger{\mathlarger{\mathfrak{a}}}} \neq 1$,
and $f_N^{\nu}$ stays non-vanishing. This prohibits conservation of $T_{N}^{\mu\nu}$. These features are precisely the ones listed in Table \ref{table1}. The MONDian force is consistent with
(\ref{einstein-mond}). Since ${\mathlarger{\mathlarger{\mathfrak{a}}}}$ is related to $\overline{g}_{\mu\nu}$ as in (\ref{relat}), the second metric
$\overline{g}_{\mu\nu}$ turns out to be a fundamental ingredient of the entire formalism. Not surprisingly, effective forces similar to $f_N^{\nu}$ also arise in
modified gravity theories which couple curvature and energy-momentum tensor $T_{N}^{\mu\nu}$ directly \cite{curve-en-mom}.

\end{enumerate}
In this section, we have succeeded to get the MONDian dynamics starting from (\ref{EFE}) by defining the acceleration scalar ${\mathlarger{\mathlarger{\mathfrak{a}}}}$ as in (\ref{relat}),
the $Q$ scalar as in (\ref{Q-scal}), and the second metric tensor as in (\ref{metric-bar}). Moreover, we have explicitly ensured conservation of the total energy-momentum tensor $T^{\mu\nu}$
while determining effective MOND force associated with the non-conservation of $T_{N}^{\mu\nu}$. The analysis here provides an existence proof.

\section{Conclusion and Future Prospects}
In the present paper, we reported our results on relativistic MOND as derived from modified dynamics rather than modified gravity. Our approach
is an empirical one and gives the beginning stage of a general investigation of relativistic MOND. The formalism developed, though lacks an action principle,
can be regarded as generalizing Milgrom's modified inertia approach \cite{modif-inertia, modif-inertia-2} to relativistic domain. It is based on
the energy-momentum tensor of matter. The reason for this is that, the energy-momentum tensor of matter in  Newtonian regime, which necessarily looses its
conservation property due to extra interactions occurring at sub-Hubble accelerations, seems to provide correct path way to quadratic acceleration in MOND regime.
In fact, this dynamical structure cannot follow from other sources  such as potentials, metric tensor and curvature tensor. The main observation behind our
approach is that, matter possesses its usual energy-momentum tensor under the usual circumstances where Newtonian laws hold. However, the same matter,
at exceedingly small accelerations below the Hubble scale, develops novel interactions causing non-conservation of its energy-momentum tensor,
and it is with these interactions that MONDian dynamics arises. Our empirical relativistic model is essentially a bi-metric theory. However,
our approach to the second metric tensor mimics models of gravitational Higgs mechanism in which the vacuum expectation value of the second
metric tensor equals the flat Minkowski metric, and it provides the requisite terms clearing the gravitational acceleration contributions to
enable the quadratic acceleration piece needed for MOND.

The present study can be extended in various aspects for rectifying and improving the present model.
\begin{itemize}
\item In the present work we have taken matter at the skirts of galaxies as dust. For an accurate analysis
of matter distribution, however, one may need to extend it to perfect fluid and other forms of matter.

\item In obtaining the MOND equation of motion (\ref{mond-eom}) we have neglected contributions from spatial variation of $\rho$.
The situation can be improved by incorporating such terms from (\ref{MOND-1}). The effect can be pronounced especially at
the arms of spirals where dust density changes sharply.
\end{itemize}
These points are currently under investigation in \cite{biz}.

Last but not least, the present model would be grossly improved if an invariant action could be written. The alleged action,
which must directly generalize Milgrom's modified inertia approach in \cite{modif-inertia} to relativistic velocities could
be too complicated to construct due mainly to the presence of the fixed acceleration scale $a_0$. It might necessitate $a_0$
to be included in relativistic transformations.

{\bf Acknowledgements.} We are grateful to Ismail Turan for fruitful discussions.

\end{document}